\documentclass[twocolumn,preprintnumbers,amsmath,amssymb,prl,showpacs]{revtex4}
\usepackage[dvips]{graphicx}
\usepackage{dcolumn}
\usepackage{bm}
\usepackage[dvips]{epsfig}
\usepackage{color}

\begin{document}

\title{Probing fluctuation-dissipation theorem in a Jean Perrin like experiment}
\author{Jean Colombani}
\author{Laure Petit}
\author{Christophe Ybert}
\author{Catherine Barentin}
\email{Catherine.Barentin@univ-lyon1.fr}

\affiliation{Laboratoire de Physique de la Mati\`ere Condens\'ee et Nanostructures;
Universit\'e de Lyon; Universit\'e Claude Bernard Lyon 1;
CNRS, UMR 5586; Domaine scientifique de la Doua, F-69622 Villeurbanne cedex, France}

\begin{abstract}
In this letter, we present a new experimental approach which allows for a direct investigation of the effective temperature concept as a generalization of the \textit{Fluctuation-Dissipation theorem} (FDT) for out-of-equilibrium systems. \textit{Simultaneous} measurements of both the diffusion coefficient and the sedimentation velocity of heavy colloids, embedded in a Laponite clay suspension, are performed with a fluorescence-recovery-based setup. This non-perturbative dual measurement, performed at a single time in a single sample, allows for a direct application of the FDT to the tracers velocity observable. It thus provides a well-defined derivation of the effective temperature in this aging colloidal gel. For a wide range of concentrations and aging times, we report no violation of the FDT in Laponite suspensions, with effective temperature agreeing bath temperature. This result is consistent with recent theoretical predictions on the coupling between velocity observable and the out-of-equilibrium gels dynamics.
\end{abstract}
\maketitle

{\it Introduction -}
The extension of equilibrium statistical mechanics tools to non-equilibrium situations is
a topic of intense activity \cite{Cugliandolo:1993, Kob:1999, Berthier:2000, Kurchan:2005,
Speck:2006, Prost:2009}.
Among these situations, the case of materials with relaxation dynamics so slow that they remain
out of equilibrium is a candidate for the extension of the Fluctuation-Dissipation Theorem (FDT).
The FDT states that the relaxation of spontaneous fluctuations and the response to a weak external
perturbation are linked by the temperature of the bath with which the system is in equilibrium \cite{Callen:1951}.
In a material exhibiting a slow structural relaxation,
whereas fast fluctuations still thermalize to the bath temperature $T_\text{bath}$, slow modes
do not, and it has been proposed that an ''effective temperature'' $T_{\text{eff}}$, greater than $T_{\text{bath}}$,
characterizes their relaxations \cite{Cugliandolo:1993}.
Its slow evolution leads the system to age, its properties to evolve and $T_{\text{eff}}$
should progressively decay toward $T_{\text{bath}}$.
As the characteristic time of slow relaxations is orders of magnitude larger than the characteristic time of fast ones,
the kinetics of both motions are clearly differentiated.
So theoretical and numerical studies have shown that the FDT should still hold for the slow structural rearrangements with
$T_{\text{bath}}$ merely replaced by $T_{\text{eff}}(t_{\text{w}},f)$, with $t_{\text{w}}$ the age of the system
and $f$ the frequency of the fluctuations and of the external force. Although some issues are still under strong debate (for instance the dependence of $T_{\text{eff}}$ on the observable \cite{Martens:2009, Russo:2010}), the concept is now well established theoretically and verified numerically.
But a clear experimental verification is still lacking.
Attempts have been made in various ageing material: structural glasses \cite{Grigera:1999}, spin glasses \cite{Herisson:2002}, polymer glasses \cite{Bellon:2002} and colloidal suspension \cite{Bellon:2002,Abou:2004,Jabbari:2007,Greinert:2006,Jop:2009,Strachan:2006, Maggi:2009}. In this last case, a large number of FDT measurements have been performed in ageing Laponite suspensions but no coherent picture emerges from these studies, where various observable have been tested and various methodologies used. On one hand, fluctuations have been evaluated by particle tracking \cite{Abou:2004,Jabbari:2007,Jop:2009}, voltage noise measurement \cite{Bellon:2002} or Dynamic
Light Scattering \cite{Strachan:2006, Maggi:2009}. On the other hand, dissipation has been evaluated by measuring the mobility of tracers optically trapped \cite{Abou:2004,Jabbari:2007,Jop:2009}, impedance \cite{Bellon:2002}, rheological properties \cite{Strachan:2006}, or by using birefringent responses \cite{Maggi:2009}. The results of these studies are extremely diverse.
Some conclude to the validity of the equilibrium FDT \cite{Jabbari:2007,Jop:2009}, whereas some others find an effective temperature greater than the bath temperature \cite{Bellon:2002, Strachan:2006, Maggi:2009,Abou:2004}.
In this last case, no consistency can be found in $T_{\text{eff}}$, neither for its order of magnitude (from $2T_\text{bath}$ \cite{Greinert:2006} to $10^3T_\text{bath}$
\cite{Bellon:2002}), nor for its evolution with the age of the gel (increasing \cite{Strachan:2006,Greinert:2006}, decreasing \cite{Bellon:2002,Maggi:2009}, or increasing and decreasing \cite{Abou:2004}). Besides the diversity of the studied observables, an other reason for these discrepancies comes from the extreme sensitivity of the Laponite gel to its elaboration, environment and ionic strength \cite{Mongondry:2005}. The settling kinetics for instance varies largely from one sample to the other, in particular it is very sensitive to the addition of colloidal probes \cite{Petit:2009}. In all the above-mentioned studies, fluctuations and dissipation are evaluated in different samples or in one sample at different times so that the resulting effective temperature does not probe a well-defined state.

In this Letter, we propose a new approach for studying the FDT inspired by the experiment of Jean Perrin on the sedimentation of heavy colloids. The fluctuating part is given by the diffusion coefficient of the colloids and the dissipative part by their sedimentation.
According to the original derivation of the FDT, the fluctuations and the dissipation are measured, in this sedimentation-based study, at one time, from one signal, in a single sample. This simultaneous measurement enables to avoid the usual pitfalls encountered in the experimental determination of $T_\text{eff}$. We find no violation of the FDT in the case of an ageing Laponite suspension. The benchmark value obtained with this method is in agreement with recent theoretical works \cite{Russo:2010}.

{\it Experiment-}
In this study, we propose an experiment inspired by the original work of Jean Perrin who measured the equilibrium concentration profile of sedimenting colloids. Indeed, we make use of the diffusion coefficient and the sedimentation velocities of heavy colloids imbedded in the Laponite suspension to assess its temperature.
However, due to the high viscosity of the gel, the equilibration is far too long compared to the ageing time of the system \cite{footnote}. But the non-equilibrium FDT is not only valid during the steady state but also during the transient regime of the sedimentation, the fluctuations of the colloids and the dissipation in the gravitational field being always thermalized at $T_\text{eff}$. Therefore, instead of investigating the final result of the gravitation-diffusion balance, we measure the sedimentation velocity and diffusion coefficient during the establishing of the concentration profile.
In its integral form, the FDT applied to the velocity of a tracer in a non-equilibrium medium
reduces to a generalized Einstein relation:
$D=\mu\,k_BT_{\text{eff}}$ where $D$ stands for the tracer diffusion coefficient, $\mu$ for the mobility in the gravitational field and $k_B$ for the Boltzmann constant.
Thereby the effective temperature can be readily deduced from the data of $D$ and $\mu$.
To perform the measurements, we use a modified version of a Fluorescent Recovery After Photobleaching (FRAP) setup giving access independently to $D$ and $\mu$ (see \cite{Petit:2009} for more details).
\begin{figure}
\includegraphics[width=0.5\textwidth]{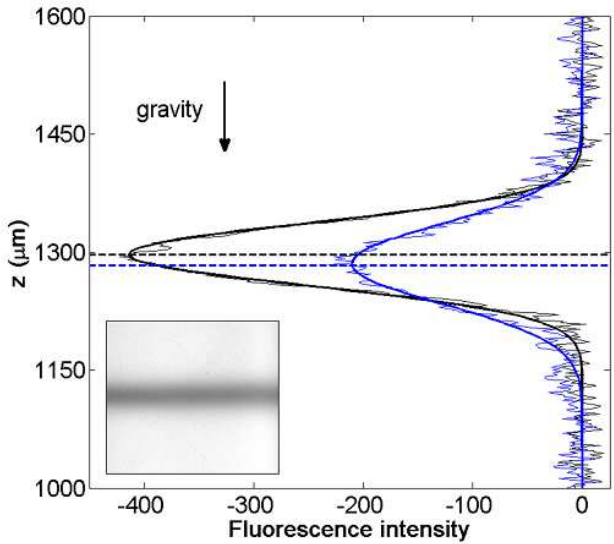}
\caption{Vertical intensity profiles emitted by fluorescent colloids in water, at time $t=8$ s (black curve) and $t=239$ s (blue curve) after a horizontal layer of 110 $\mu$m thickness has been photobleached. \textcolor{red}{Insert: Picture ($0.8\times 0.9$\,mm) of the photobleached slab}.
The gaussian fits of the two intensity profiles are also plotted.
\label{nappe}}
\end{figure}
In this technique, a Laponite RD sol with concentrations ranging from 1.6\%wt to 2.2\%wt is \textcolor{red}{uniformly} seeded with silica fluorescent colloids of diameter $d=200$~nm and weight density $\rho =1.95\times 10^3$ kg m$^{-3}$ (manufactured by FluoProbes$^{\text{\textcircled{\tiny R}}}$), the volume fraction of the fluorescent tracers being $0.1\%$.
This sol is then placed in a $0.25\times 0.25\times 20$ mm$^3$ optical cell closed by polytetrafluoroethylene seals to prevent evaporation.
Zero time corresponds to the filling of the cell.
Then, a horizontal layer of the sample with a Gaussian vertical profile is bleached by a 2 W green Argon laser.
Through time, the Gaussian intensity of the photobleached layer, characterized by its width $w$ and its maximum position $z_0$, broaden due to the colloids diffusion and goes downward under colloids sedimentation, as shown in Fig. \ref{nappe}. Note that broadening and position shift are typically of the order of $10$ to $50$ $\mu$m.
\begin{figure}
\begin{tabular}{@{}c@{} @{}c@{}}
\includegraphics[width=.5\linewidth]{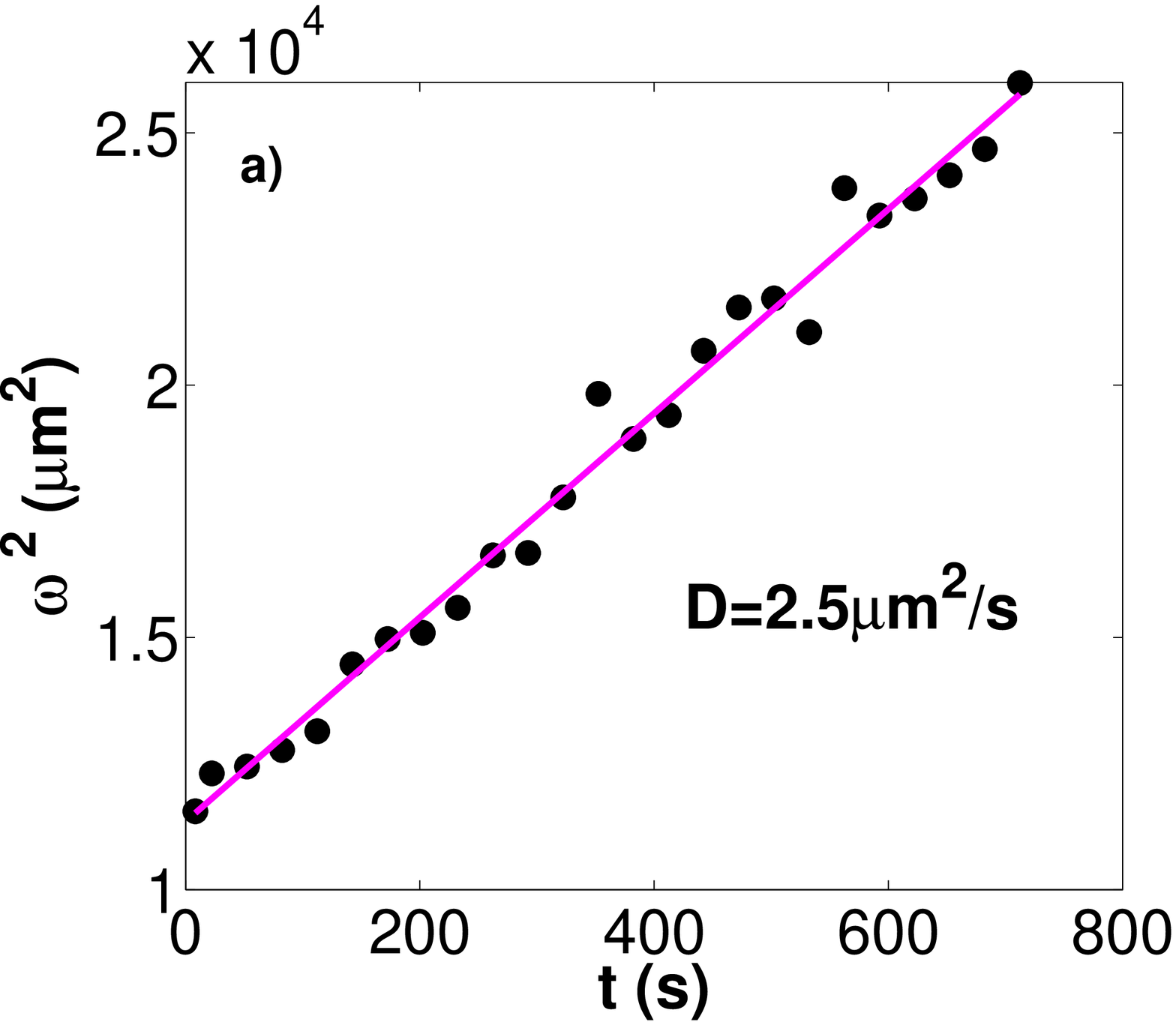} & \includegraphics[width=.5\linewidth]{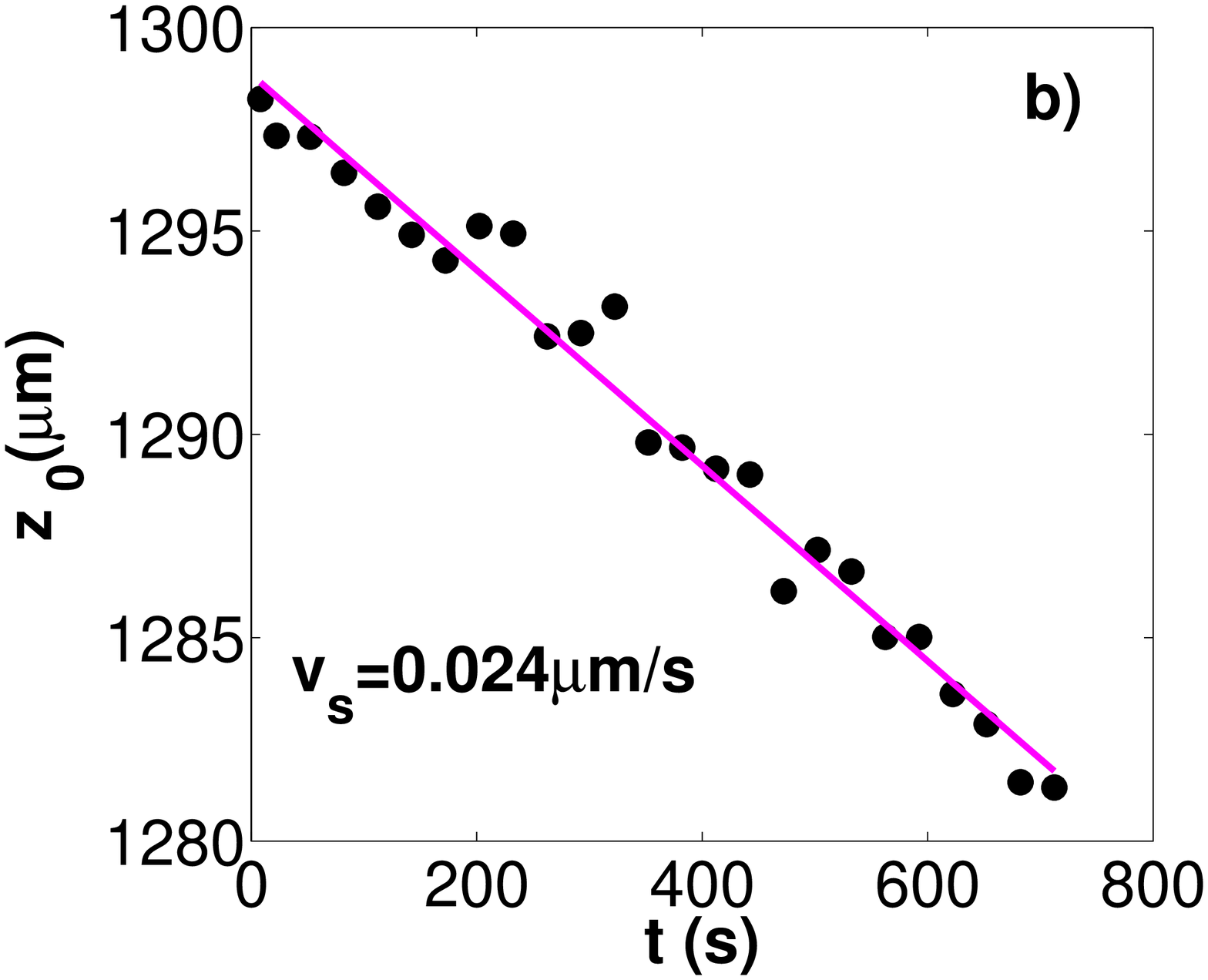}
\end{tabular}
\caption{Evolution with time, for one experiment, of a) the squared width and b) the position of the mean of the Gaussian intensity profile allowing to assess the diffusion coefficient $D$ and the sedimentation velocity $v_\text{s}$ of the tracers.}
\label{width_mean}
\end{figure}
In Fig. \ref{width_mean} are plotted the evolutions with time of $w^2$ and $z_0$, which both increase linearly with time. The associated slopes yield respectively the diffusion coefficient $D$ and the sedimentation velocity $v_{\text{s}}$ of the colloids. The temperature of the sol is directly deduced from those measurements and is given by the following relation:
\begin{equation}
T_{\text{eff}}=\frac{D}{v_\text{s}}\frac{\Delta\rho\,g\,\pi\,d^3}{6\,k_B}
\end{equation}
where $\Delta\rho$ is the difference of densities between the silica beads and the sol and $g$ the acceleration of the gravity.

{\it Validation of the FRAP technique -}
The Laponite sol is a suspension of charged disk-shaped beads (30 nm diameter, 1 nm thickness) of a synthetic clay dispersed in water, with typical mass fraction ranging from 1 to 4 \%wt. In this concentration range, the nature of the resulting solid  (repulsive glass of Debye spheres or attractive gel with percolating agregates) remains under discussion \cite{Ruzicka:2004}
but whatever the type of arrested state, the Laponite sol is characterized by a loose structure. Molecular tracers diffuse in Laponite suspension as in water without hindrance of the platelets. Tracer diffusion starts to be affected by the Laponite network when the probe diameter exceeds 30 nm, which is the typical mean distance between the platelets \cite{Petit:2009}. Therefore we have chosen the colloids diameter larger than this characteristic length, in the aim to probe the gel and not merely the solvent.
In our experiment, the Peclet number defined by $\text{Pe}=v_\text{s}\, L/D$, \textcolor{red}{$L$ being the typical sedimentation length},
is always of the order of 1 which guarantees a comparable magnitude of diffusion and sedimentation. The time window ranges from 100 to 500 s, which is always much smaller than the settling time, ranging from 6h to 2 days, depending on the concentration of the Laponite suspensions.
In these conditions, the age of the investigated sample remains well defined.
For this reason Laponite concentration was limited to 2.2\%wt to avoid its too fast solidification.
Note that measuring the dissipation through sedimentation is as non-intrusive as possible.
Indeed, the external force due to the gravitational field exerted on each colloidal probe is about $4\times 10^{-5}$ pN, which is typically 4 or 5 orders of magnitude smaller than the forces exerted on colloids by optical traps \cite{Jop:2009}. In terms of stress, the sedimentation of the colloids induces a stress of $~0.3$ mPa, much below any reported yield stress \cite{Willenbacher:1996}. The only disturbing influence of the technique is the possible heating by the photobleaching process
of the fluorescent colloids, their quantum yield being roughly 50\%, and of the solvent, the absorption coefficient of water being  minimum but nonzero at this wavelength.
Using the energy of the photobleaching laser sheet, the colloids concentration and quantum yield, the solvent absorption coefficient and heat capacity,
the temperature increase can be estimated as less than 0.1 K, therefore assumed to be of negligible influence.
\begin{figure}
\includegraphics[width=\linewidth]{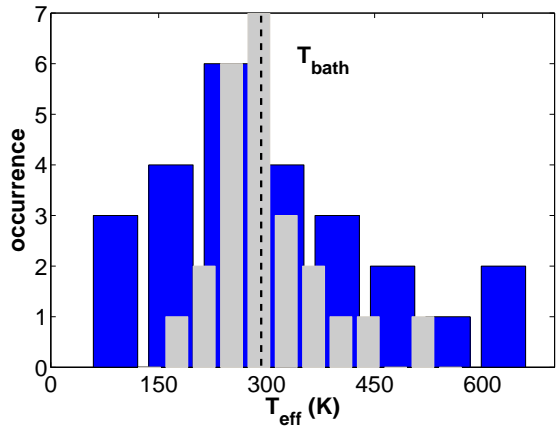}
\caption{Histograms of the temperatures measured in water (in grey) and of those measured in the Laponite suspensions (in blue).
\label{histotout}}
\end{figure}

Finally, to assess the reliability of our methodology, a number of experiments has been carried out in ultrapure water.
Various positions in the cell, laser powers, photobleaching times, waiting times since the filling of the cell, volume fraction of the probes, and time windows were tested, to estimate the variability of the method.
All values are gathered in the histogram of Fig. \ref{histotout}.
The fit of the histogram with a normal law brings $T_{\text{eff}}=(294 \pm 14)$ K.
The standard error is obtained by $1.06\times\sigma/\sqrt{N}$, $\sigma$ being the standard deviation and 1.06 the Student coefficient for $N=24$ measurements and a 70\% confidence interval.
This result enables to appreciate the rightness of our thermometer, bringing exactly the bath temperature of our experiment (293 K). It also provides the accuracy of the measurement, i.e., $\simeq5$\% in a simple fast relaxing system.
\begin{figure}
\includegraphics[width=\linewidth]{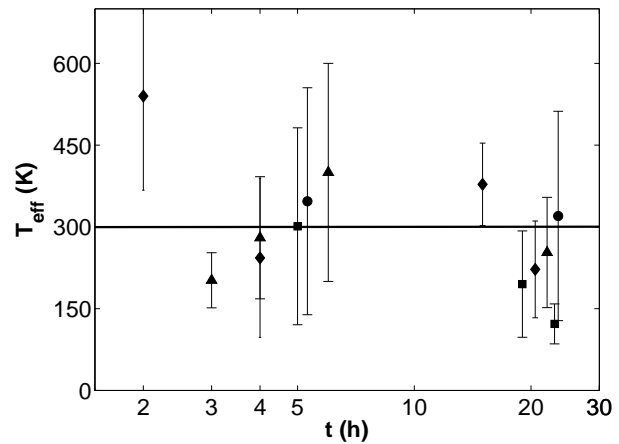}
\caption{Evolution with time of the effective temperatures measured in Laponite suspensions of different concentrations (symbol, concentration in $\%
$wt): ($\blacktriangle$, 1.6), ($\blacklozenge$, 1.8), ($\blacksquare$, 2), ($\bullet$, 2.2).
Each dot corresponds to the average of roughly 3 consecutive measurements in the same sample. The error bar corresponds to the standard deviation.
\label{Tefflap_t}}
\end{figure}

{\it Results and Discussion-}
Measurements of the diffusion coefficients and sedimentation velocities of colloids have been performed along the same protocol in Laponite suspensions, from which the effective temperature has been calculated.
As in water, various experimental parameters have been tested.
\textcolor{red}{The sensitivity of the device has enabled to investigate rather old suspensions, from 1.5 to 24 hours, compared to standard techniques \cite{Jabbari:2007,Jop:2009,Abou:2004,Strachan:2006,Greinert:2006}.}
Fig. \ref{Tefflap_t} displays the obtained temperature as a function of the age of the samples for various concentrations.
\textcolor{red}{All values are below 1.8$T_\text{bath}$ and after 3h are all distributed in the interval $T_\text{bath}\pm 165$\,K.
Although this statistical dispersion is large, it is of the same order of magnitude as all reported uncertainties \cite{Abou:2004,Bellon:2002,Greinert:2006,Jabbari:2007,Jop:2009,Strachan:2006}.}
Second no significative variation of the temperature with the concentration of the sample can be observed.
Third despite the dispersion of the results, no evolution with the age of the gel is discernable.
This constancy of $T_{\text{eff}}$ derives from the simultaneous decrease of the diffusion coefficient (from $3$ $\mu$m$^2/$s to $0.03$ $\mu$m$^2/$s) and settling velocity (from $30$ nm$/$s to $0.3$ nm$/$s) during the solidification of the suspension.
For samples older than roughly one day, the values of these quantities go below the sensitivity of our device.
As the effective temperature measured in all samples does not seem to exhibit any characteristic evolution with the concentration or with the age of the Laponite, an overall histogram of all the measurements has been constructed (Fig. \ref{histotout}).
A normal law fitted to the histogram obtained in Laponite gives $T_{\text{eff}}=(305 \pm 33)$ K for 25 measurements.
\textcolor{red}{Our measured standard deviation indicates that our technique is not able to discern deviation from equilibrium smaller than 165 K.
But the statistical analysis shows that the laponite suspension effective temperature lies within 10\% of the bath temperature.}
Our experimental results support recent numerical investigations by Sciortino \& Russo
\cite{Russo:2010}, who show that the velocity observable relaxes fast, due to the loose nature of the setting gel, and therefore does not probe the slow relaxation
of the arrested state.
Thereby, a consistent picture of the out-of-equilibrium state of colloidal gels appears, with no violation of the FDT for fast-relaxing observable,
like velocity, and an effective temperature decreasing with time to bath temperature for observable sensitive to slow relaxing modes, like electric charge
\cite{Bellon:2002} or orientation of the Laponite platelets \cite{Maggi:2009}.
Beside we can interpret this absence of violation of the FDT by the fact that the measurements of tracer diffusion are not coupled to the dynamical heterogeneities
of the system, so that they are not sensitive to its out-of-equilibrium nature \cite{Berthier:2011}.


{\it Conclusions-}
In this Letter, using a fluorescence recovery technique, we measure simultaneously the velocity fluctuations and the mobility in the gravitational field of tracers imbedded in ageing Laponite gels. Therefore the relaxation of the tracers to spontaneous fluctuations and to an external force are probed jointly, in strictly the same environment, according to the original idea of
the fluctuation-dissipation theorem.
Using this procedure which guarantees an unprecedented validity of effective temperature measurements,
we find no violation of the FDT for the velocity observable.
This feature finds an explanation in a recent numerical work.
Under the light of this study and of our experiments, a critical analysis of the literature results enables to build a coherent picture of $T_\text{eff}$
measurements in gels, where the velocity observable thermalizes to the bath temperature whereas other observable, like the electric current or platelet orientation, do not.
Now that benchmark results for the FDT in a soft glass at rest have been obtained, next work will focus on the FDT behavior in a soft glass under shear.
In this situation, either the imposed flow induces a rejuvenation of the material and an increase of its effective temperature, or an overaging and a decrease of the effective temperature remains an open question \cite{Berthier:2002}.


\begin{acknowledgments}
The authors wish to thank L.~Bocquet, J.L.~Barrat, P.~Jop and S. Jabbari-Farouji for fruitful discussions.
This work was supported by Agence Nationale de la Recherche under project SLLOCDYN.
\end{acknowledgments}


\begin{thebibliography}{}
\bibitem{footnote} For a typical viscosity of 1 Pa s, a typical size of silica tracers of $0.1$ $\mu$m, the sedimentation time on a distance of $100$ $\mu$m is of the order of 1000 h.
\end{thebibliography}
\end{document}